# Giving electrons a ride: nanomechanical electron shuttles

Andriy V. Moskalenko[1], Sergey N. Gordeev[1]*, Olivia F. Koentjoro[2], Paul R. Raithby[2], Robert W. French[2], Frank Marken[2] and Sergey E. Savel'ev[3]

[1] Department of Physics, University of Bath, Claverton Down Road, Bath, BA2 7AY, UK

[2] Department of Chemistry, University of Bath, Claverton Down, Bath, BA2 7AY, UK

[3] Department of Physics, Loughborough University, Loughborough LE11 3TU,

*Corresponding author. E-mail: s.gordeev@bath.ac.uk

Abstract

Nanomechanical shuttles transferring small groups of electrons or even individual electrons from one electrode to another offer a novel approach to the problem of controlled charge transport. Here, we report the fabrication of shuttle-junctions consisting of a 20 nm diameter gold nanoparticle embedded within the gap between two gold electrodes. The nanoparticle is attached to the electrodes through a monolayer of "flexible" organic molecules which play the role of "springs" so that when a sufficient voltage bias is applied, the nanoparticle starts to oscillate transferring electrons from one electrode to the other. Current-voltage characteristics for the fabricated devices have been measured and compared with the results of our computer simulations.

In 1998, a group of theoreticians[1] proposed a novel mechanism of charge transport in nanostructures based on mechanical shuttling of electrons. The basic element (see Fig. 1), which is called a shuttle-junction, consists of a metallic nanoparticle connected by flexible, "elastic" molecules to two nanoelectrodes. For low applied voltages where the nanoparticle is stationary the device is similar to a single-electron transistor (SET).[2] The vibrational process can be initiated by applying a sufficiently large bias voltage to the electrodes. This will induce the tunnelling of electrons from the negative electrode to the metallic particle. Because of the Coulomb blockade effect, only a limited number $N$ of electrons will be loaded onto the particle, with $N$ being dependent on the volume of the particle and the applied voltage. Then, under the influence of the electrostatic field applied between the electrodes, the particle will move to the positive electrode bringing with it the charge $-Ne$. Near the positive electrode the probability of tunnelling increases and $2Ne$ electrons from the nanoparticle jump onto the positive electrode. This results in the deficit of electrons and the nanoparticle loaded by a positive charge, $+Ne$, starts to move back to the negative electrode to load a new bunch of electrons. Such charging-recharging processes can be repeated again and again. This cyclic "electron shuttle" mechanism of discrete charge transfer gives rise to a current $I=2Nef$ through the nanostructure, which is proportional to the vibration frequency $f$.

One of the potential advantages of the shuttle-junctions is that, in contrast to SETs, *only one tunnelling barrier is open at a time.*[1] This prevents simultaneous tunnelling of two electrons trough the barriers (so-called co-tunnelling effect[3-4]) thus increasing accuracy of the single-electron transport. Secondly, while in a SET the electric current is controlled by the rate of the electron tunnelling between the immobile island and electrodes, in the shuttle-junction the current is expected to be determined by the vibration frequency of the nanoparticle. The reason for this is that when the electrons jump to or from the electrode, the nanoparticle is so close that the tunnelling rate is large compared to the shuttle vibration frequency which can be as high as $10^{11}$ Hz.[1] This should lead to a significant increase of the current through the device in comparison with the stationary island. Exploitation of this mechanism could potentially lead to the development of a new generation of nanomechanical electronic devices, such as transistors, current standards, very sensitive electrometers, sensors, logic gates, and memories with ultra low power consumption and high speed of operation.

Such shuttles at certain conditions should enable coupling between electromagnetic and mechanical degrees of freedom (for instance via excitation of plasmons in metallic particles).

Interest in electron shuttles and, more generally, in the fundamental properties of electromechanical coupling in nanostructures, initially generated by theoretical papers,[1, 5-6] (see also Ref. 7 for a review), has grown tremendously over the last few years as experimentalists made the first attempts at the practical realisation of such devices. There were reports about manufacturing of different nanoelectromechanical systems (NEMS) using semiconductor fabrication technology. In these systems the charge transfer was performed by a flexible semiconductor element such as a silicon beam resonator,[8] clamped GaAs beam,[9] cantilever,[10] silicon nanopillar [11] *etc*. However, the functional elements of all these devices are comparatively large (larger than 100 nm), work at a fixed frequency and usually required an additional ac driving voltage to set up oscillations of the flexible element at its resonant frequency. Very elegant experiments have been performed by Park et al. [12] who measured electron transport through a molecule of $C_{60}$ that was trapped between two electrodes during the electromigration process. However, as $C_{60}$ molecules are only ~1 nm in diameter, it was impossible to take an image that would confirm trapping of the molecule inside the nanogap. The observed current-voltage characteristics can be interpreted in terms of shuttling, but alternative explanations have also been suggested.[13-14]

In the present work, we describe the fabrication and properties of the first metallic electromechanical nanoshuttles consisting of a 20 nm gold nanoparticle embedded in the gap between two electrodes and attached to them through a monolayer of flexible organic molecules. We have found that the geometry of shuttle-junctions suggested by Gorelik *et al.* [1] (Fig. 1) is very inconvenient from the point of view of fabrication as it requires a high-precision matching between the size of the gap and the shuttle diameter. To reduce requirements to accuracy of fabrication we have elaborated a structure which is depicted schematically in Fig. 1c. We fabricated the nanoelectrodes with rounded edges so that small variations in the nanoparticle diameter are less critical than in Fig. 1a.

Shuttle-junction devices were formed on the top of a silicon wafer coated with a ~1 μm $SiO_2$ layer. Planar 30-nm thick gold nanoelectrodes separated by a gap of 10-20 nm were fabricated using electron-beam lithography followed by lift-off. When parameters of exposure, metal deposition and the lift-off process are optimised, this technique gives smooth electrodes with rounded edges required for the design shown in Fig. 1c. Figs 2a and 2b show an example of nanoelectrodes used for fabrication of our shuttle-junctions. The images were taken using a VEECO Multimode IIIa Atomic Force Microscope with WSxM software [15] for making pseudo-3D AFM images.

To assemble shuttle-junctions, we initially covered the nanoelectrodes with a monolayer of 1,8-octanedithiol which have a length of approximately 1.2 nm. Then nominal 20 nm diameter gold nanoparticles (G1652 Sigma-Aldrich) were adsorbed by immersion of the electrode assembly into aqueous gold sol. An AFM image of the area around the nanogap was taken and one of the nanoparticles was manipulated into the gap using the AFM tip. Fig. 2c shows a sequence of AFM images obtained during such a manipulation. Fig. 2d-f shows AFM images of some of the shuttle-junctions. Current-voltage characteristics of the fabricated devices were measured at room temperature in a shielded dry box to protect samples from moisture and to decrease electromagnetic noise.

For simulations of the nanoparticle dynamics we used the model proposed in Refs 1, 16, where, in addition, we took into account that the nanoparticle could be pinned by a potential produced during nanofabrication: $\alpha d^2x/dt^2 + dx/dt + \omega_0^2 x = nu + 4n^2x - U'$. Here the first term describes the particle inertia with dimensionless mass $\alpha$, the second term is related to the viscous force, and the third term is associated with the elastic force having the dimensionless rigidity $\omega_0^2$. The right-hand-side describes both the electrostatic force acting on the nanoparticle carrying $n$ electrons (where the dimensionless voltage $u$ is normalized by the Coulomb blockade threshold $V_c$) and the pinning force which is described by the dimensionless potential $U$ with the prime denoting a spatial derivative. We used the simplest parabolic potential, i.e., $U'=F_c x/d$ for $|x|<d$ and 0 otherwise. Here, $F_c$ is the maximum pinning force and $d$ is the size of the pinning trap. The position of the particle $x$ (the central position corresponds to $x=0$) is normalized by $2L$ where $L \sim 1$nm is the size of the gap between particle and an electrode, while the time $t$ is normalized by a characteristic time $t_0$ which is determined by viscosity. Electron tunneling is determined by the probabilities $P^{\pm}_{L,R}$ to jump from either the left (L) or right (R) junction towards either the right (+) or the left (-) during time $\Delta t$: $P^{\pm}_{L,R} = \Delta t \Delta G^{\pm}_{L,R} / \{e^2 R^{\pm}_{L,R}[1-\exp(-\Delta G^{\pm}_{L,R}/E_T)]\}$, where $\Delta G^{\pm}_{L,R}$ is the decrease of the free energy in the system as an electron tunnels; tunneling resistance $R^{\pm}_{L,R}$ obeys exponential law $R_L(x)=R_R(-x)=R_0 \exp(xL/\lambda)$ with a constant $R_0$ (which was taken 1TOhm) and the tunneling length $\lambda$ is less or about 0.5Å, and $E_T = k_B T$ is the thermal energy with T=300K. It is important to stress that the energy change $\Delta G^{\pm}_{L,R}$ depends on the particle position as $\Delta G^{\pm}_{L,R} = 0.5\, e\, V_c (1 - 4x^2) [\pm 2jn \pm u/(1+2jx)]$ with j=±1 for (R/L) indices. At first glance, these equations contain many adjustable parameters. However, we can estimate most of them. Indeed, $V_c$ is fixed by the capacitance of the shuttle and is about 0.16V. Parameter $\alpha$ is determined by the shuttle mass $M = 8\times10^{-20}$ kg: $\alpha = L^2 M/eV_c t_0^2 = 3/t_0^2$ if $t_0$ measured in nanoseconds. As measured in Ref. 17, the elastic constant for one molecule is about $5\times10^{-3}$ N/m. Assuming that shuttle is linked by 10 molecules we estimate the elastic constant $K$ as $5\times10^{-2}$ N/m, resulting in $\omega_0=(KL^2/eV_c)^{1/2}=1.4$. So we have only three fitting parameters $F_c$, $d$, and $t_0$. We fixed $t_0$=2000 ns for all simulations, and assumed almost point-like pinning $(F_c, d) = (150, 10^{-5})$, $(95, 4\times10^{-5})$, $(130, 4\times10^{-5})$ for Fig.4 (top, middle, and bottom curve, respectively). Small values of $d$ indicate the small probability of re-trapping of nano-shuttle by the pinning site.

Fig. 3 shows experimental I-V curves for three fabricated shuttle-junctions compared with results of our computer simulations. It can be seen that the experimental curves show an abrupt rise in the current but at significantly higher voltages than predicted (0.16 V) by the theory [1] (dotted line). This delay in the current rise can be attributed to the effect of pinning which was not taken into account in the original theoretical model.[1] Different interactions such as van der Waals, adhesion and Casimir forces may result in a quite strong locking of the nanoparticle in a particular position within the gap. Thus, at low applied voltages the nanoparticle is stationary as the electrostatic forces between the nanoparticle and the electrodes are too small to depin it. We have added a pinning potential to the theoretical model [1] and have obtained I-V curves (dashed lines) that give much better fit to our experimental results. At low voltages, the shuttle is in the pinned state and the electrical current through the device is due to sequential tunnelling of electrons from the source electrode to the nanoparticle and then to the drain electrode. The shuttle-junction in this regime is similar to a double tunnel junction. As the voltage is increased the electrostatic force applied to the nanoparticle will grow, firstly, because more electrons will reside on the nanoparticle and, secondly, because larger electric field will be applied to it. Eventually a voltage $V=V_{depin}$ will be reached where the nanoparticle can escape its locked position and start to vibrate (see an animation of this scenario on the website http://www.lboro.ac.uk/departments/ph/research/animations/Shuttle1.swf). The amplitude of vibrations will increase until a balance is achieved between dissipated and adsorbed energies. The number of electrons transferred in one cycle will depend on the capacitance, $C$, of the shuttle, its vibration frequency, and applied voltage, $V$. Transition to the shuttling regime manifests itself as a sudden increase in the transmitted current. So large change in the current through a shuttle-junction can be associated with the so-called hard excitation of oscillations,[16] which is typical for system with large dissipation.

In order to prove that the measured current in our experiments was transferred through the nanoparticle rather than being caused by leakage, we have performed several tests. First of all, we checked that octanedithiols have a high leakage resistance. It is known that results obtained for conductance of organic monolayers strongly depend on geometry of the probes and environment.[18-19] So we tried to keep these parameters the same as in our studies of shuttle-junctions and prepared test-samples by depositing from solution 20 nm gold nanoparticles on an Au <111> surface, preliminary coated with a monolayer of octanedithiols. Then, using the tip of a Scanning Tunnelling Microscope, we applied voltage between individual nanoparticles and the Au <111> surface and measured current through the layer of molecules. The observed currents did not exceed 0.25 nA at 5V (dash-dotted line and lower inset in Fig. 3), which is much smaller than the current transferred by shuttles. We also compared characteristics of a working shuttle-device (curve 1 in Fig. 4) with the same device but from which we carefully removed the nanoparticles (curve 2 in Fig. 4). After the nanoparticle was removed, current through the device dropped to the level of noise.

The maximum number of electrons carried by a nano-shuttle is determined by the condition $N_{max}=[CV/e+0.5]$.[1] For larger $N$, the addition of an extra electron becomes energetically unfavourable. For our gold spherical particles with radius R=10nm, we estimate self-capacitance as $C = 4\pi\varepsilon_0 R \sim 10^{-18}$ F, which provides $N_{max}$ of about 20 electrons for the applied voltage of ~3 V. However, if damping is strong then the shuttle may carry fewer electrons as it will not be able too approach an electrode close enough to get the maximum load. The actual number of electrons sitting on the nano-shuttle will fluctuate within a certain range ($N-\Delta N$, $N+\Delta N$) thus producing fluctuations in the average electrostatic force applied to the particle. The shuttle starts to oscillate as soon as the electrostatic force for $N+\Delta N$ electrons exceeds the pinning force. However, the nanoparticle can be re-trapped by the pinning site if the occupation number reduces (e.g., $N-\Delta N$) due to fluctuations. Such events produce occasional breaks in shuttling and the associated sudden steps in the *I-V* curves, which are clearly seen in both experimental and simulated curves. Three simulation curves which were obtained for slightly different values of the pinning force and the size of the pinning trap are presented in Fig. 3. As can be seen, these small variations in the pinning parameters have significant effect on the shape of the *I-V* curves and the number of the steps observed. Two upper insets in Fig. 3 show shuttle displacement as a function of time at points below and above the depinning threshold, at *V*=3V, for one of the simulated *I-V* curves. Below the threshold, the nanoparticle is making rare attempts to oscillate when the fluctuating number of electrons accumulated on the particle becomes large enough. However, it becomes re-trapped again as soon as the number of electrons sitting on it fluctuates to a lower value. This process, a "precursor" to the depinning, produces a small current below 3V. Above the threshold the nanoparticle oscillates but the oscillations are quite irregular and, importantly, the amplitude of displacement of the nanoparticle varies significantly. The nanoparticle rarely approaches the electrodes closely enough to gain the maximum charge ($N_{max}$=20 for 3V) and on average it carries only 4-7 electrons. Such behaviour, when a nanoparticle spends some time trapped by a pinning potential and then some time oscillating between the electrodes and transporting electrons, is typical for the pinning-controlled regime.

In conclusion, we fabricated shuttle-junctions consisting of a 20 nm gold nanoparticle attached to two electrodes through a monolayer of flexible organic molecules. Measured current-voltage characteristics have been compared with results of computer simulations and found to be in correspondence with the shuttling mechanism of charge transport.

**Acknowledgements.** This work was funded by the EPSRC under grant GR/S47793/01 and partially supported by EPSRC via EP/D072581/1. R.W.F. thanks the Innovative electronic Manufacturing Research Council (IeMRC) for a studentship (no. GR/T07549). SS acknowledges DML, RIKEN for access to the computer facilities.
**References**

1. L.Y. Gorelik *et al.*, Phys. Rev. Lett. **80**, 4526 (1998).
2. K.K. Likharev. IEEE Trans. Mag. **23**, 1142 (1987).
3. D.V. Averin, Y.V. Nazarov, Phys. Rev. Lett. **65**, 2446 (1990).

4. T.M. Eiles *et al.*,. Phys. Rev. Lett. **69**, 148 (1992).

5. L.Y. Gorelik *et al.*, Nature **411,** 454 (2001).

6. A.D. Armour, A. MacKinnon. Phys. Rev. *B* **66**, 035333 (2002).

7. R.I. Shekhter *et al.*, Phys.: Condens. Matter **15**, R441 (2003).

8. R.H. Blick *et al.*, Physica E (Amsterdam) **6**, 821 (2000).

9. R.G. Knobel, and A.N. Cleland, Nature **424**, 291 (2003).

10. A. Erbe *et al.,* Phys. Rev. Lett. **87**, 096106 (2001).

11. D.V. Scheibe, and R.H. Blick, Appl. Phys. Lett. **84**, 4632 (2004).

12. H. Park *et al.* Nature **407**, 57 (2000).

13. D. Boese, and H. Schoeller, Europhys. Lett. **54**, 668 (2001).

14. K.D. McCarthy *et al.*, Phys. Rev. *B* **67**, 245415 (2003).

15. I. Horcas *et al.*, A. M. *Rev. Sci. Instrum.* **78**, 013705 (2007).

16. T. Nord *et al.*, Phys. Rev. B **65**, 165312 (2002).

17. Y. Sakai *et al.*, Appl. Phys. Lett. **81**, 724 (2002).

18. X.D. Cu *et al*., Science **294**, 571 (2001).

19. D. Long *et al.*, Nature Materials **5**, 901 (2006).

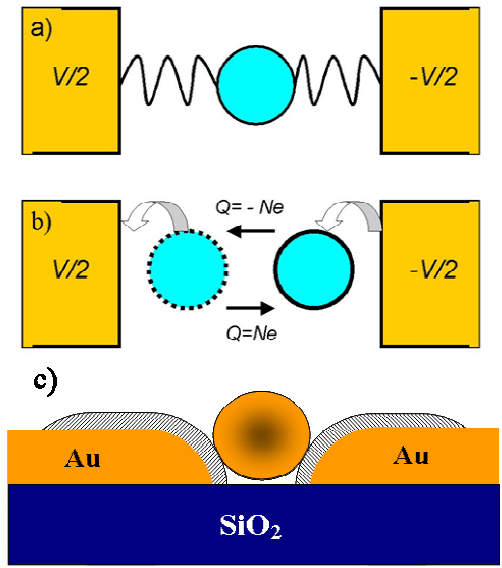

FIG. 1. Theoretical model of an idealised shuttle-junction (a), illustration of the shuttling process (b) [Ref. 1] and (c) experimental realisation of a shuttle-junction. The device consists of a 20 nm gold nanoparticle attached to two gold electrodes through monolayers of octanedithiol molecules serving as springs.

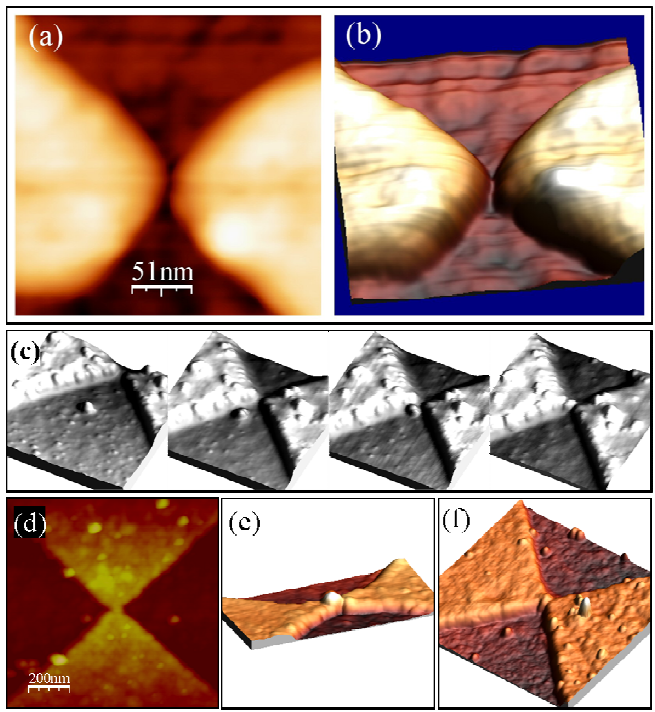

FIG. 2. Assembly of shuttle-junctions.
(a) and (b) are, respectively, the top view and pseudo-3D images of electrodes used for fabrication of shuttle-junctions; (c) sequence of AFM images taken during manipulation of a 20-nm nanoparticle into the gap between two electrodes; (d)-(f) AFM images of fabricated shuttle-junctions.

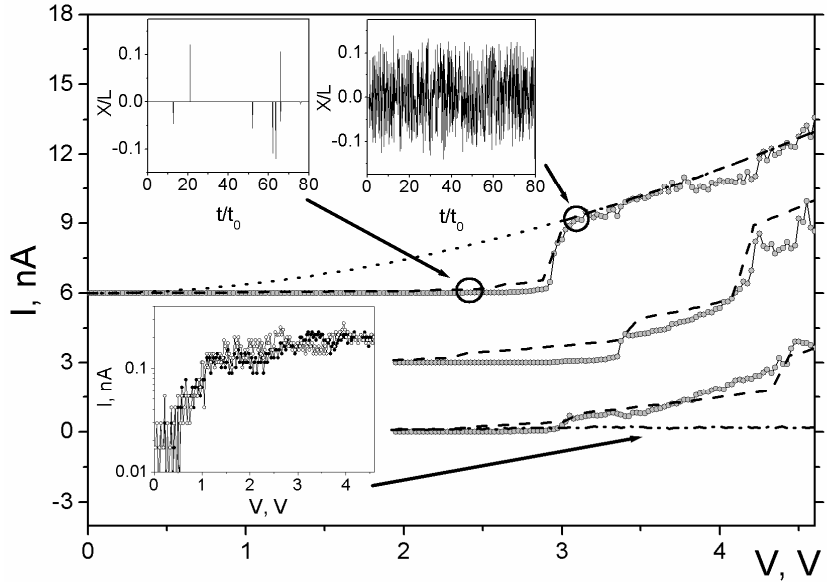

FIG. 3. Experimental (symbols) and simulated (dotted and dashed lines) current-voltage characteristics for shuttle-junctions fabricated in the present work. The dotted line corresponds to oscillations in the case of zero pinning and the dashed lines to the case of finite pinning in the system (see text). The two top insets show the shuttle displacement as a function of time for two points, one of which is below and the other is above the transition into the shuttling regime. The bottom inset and dash-dotted curve show the leakage current through a monolayer of octanedithiol molecules.

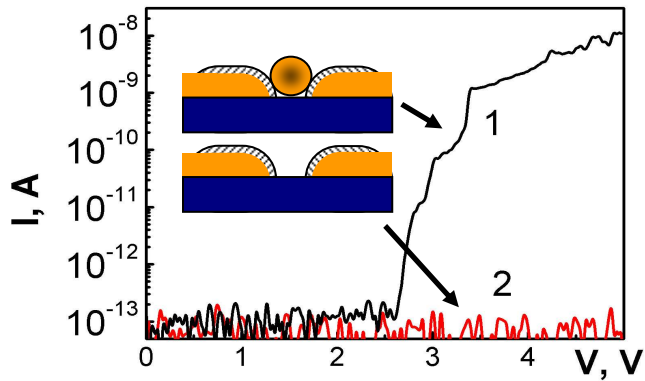

FIG. 4. To prove that current does flow through the nanoparticle in the shuttling regime (curve 1) we removed the nanoparticle from the gap using the AFM tip. This resulted in a drop of the current through the device of several orders of magnitude (curve 2). Insets demonstrate schematically the sample geometry corresponding to the curves 1 and 2.